# Separability of 3-qubits density matrices, related to $l_1$ and $l_2$ norms and to unfolding of tensors into matrices

Y. Ben-Aryeh [*] and A. Mann [†]

*Physics Department, Technion-Israel Institute of Technology, Haifa 32000, Israel*

[*] $phr65yb@physics.technion.ac.il$ ; [†] $ady@physics.technion.ac.il$

We treat 3-qubits states with maximally disordered subsystems (MDS), by using Hilbert-Schmidt (HS) decompositions, where in the general case these density matrices include 27 HS parameters. By using "unfolding methods", the MDS tensors are converted into matrices and by applying singular values decompositions (SVD) to these matrices the number of the HS parameters, in the general MDS case, is reduced to 9 and under the condition that the sum of absolute values of these parameters (the $l_1$ norm) is not larger than 1, we conclude that the density matrix is fully separable and we get an explicitly separable form for it. In another method we divide the 27 HS parameters into 9 triads where for each triad we calculate the Frobenius ($l_2$) norm of 3 HS parameters. If the sum of nine $l_2$ norms is not larger than 1 then we conclude that the density matrix is fully separable and we have another explicitly separable form for it. The condition for biseparability of MDS density matrices is obtained by the use of one qubit density matrix multiplied by Bell entangled states of the other two qubits. By using this method, the 27 HS parameters are divided into 9 triads which are different from those used for full separability. If the sum of the nine $l_2$ norms for these different triads is not larger than 1, we conclude that the density matrix is biseparable. We demonstrate the use of our methods in examples comparing the condition for full separability with the improved conditions for biseparability. We analyze the relations between 3 qubits MDS density matrices and the method of high order singular value decomposition (HOSVD) and show that this method may improve the sufficient condition for full separability. We demonstrate the application of this method in two examples. For 3-qubits states which are non-MDS the HS decomposition includes up to 64 parameters. If the sum of the absolute values of the non-MDS HS parameters is not larger than 1, we may conclude that the density matrix is fully separable, and we have explicit expressions for their separability. For the systems of GHZ state mixed with white noise and W state mixed with white noise we find a simple way to reduce the sum of the absolute values of the HS parameters absolute values and get better conditions for their full separability.

## I. Introduction

There is much interest in quantum entangled states due to various potential applications that use quantum properties of such states. The most famous application is the use of quantum systems for a new generation of computers that will be based on principles of quantum computation (QC). The building blocks of QC are usually taken as combinations of qubits states where qubit is defined as a quantum two-level system, and there are many physical systems representing qubits. A fundamental theoretical issue in this field is the distinction between states which are separable and those which are entangled.

The definition of separability of a bipartite system is: A density state $\rho$ on Hilbert space $H_A \otimes H_B$ where $A$ and $B$ are the two parts of a bipartite system is defined as *non-entangled/separable* if there exist density operators $\rho_A^{(j)}, \rho_B^{(j)}$ and $p_j \geq 0$ with $\sum_j p_j = 1$ such that

$$\rho = \sum_j p_j \rho_A^{(j)} \otimes \rho_B^{(j)} \quad . \tag{1}$$

The interpretation for such definition is that for bipartite separable states, these states are completely independent of each other. We use the following generalized definition of full-separability for a system composed of three parts A, B, C:

$$\rho = \sum_j p_j \rho_A^{(j)} \otimes \rho_B^{(j)} \otimes \rho_C^{(j)} \quad . \tag{2}$$

A similar definition can be given for larger n-qubits systems ($n > 3$).

In our previous works [1-4] we have treated extensively the separability and entanglement properties of two-qubits and three qubits systems by the use of Hilbert-Schmidt (HS) decompositions. The outer products in the HS decompositions are of Pauli matrices where we relate the correlations of 2-qubits systems to one and two qubits measurements, and the correlations for 3-qubits systems to certain one, two and three-qubits measurements, etc. One advantage of this method is that it is valid for both pure and mixed states. In a previous work [1] we analyzed separability and entanglement properties of 3-qubits systems with maximally disordered subsystems (MDS) [5], i.e., density matrices which, by tracing over any subsystem, will give the unit density matrix, e.g.

$$Tr_A \rho(A,B,C) = (I)_B \otimes (I)_C / 4 \qquad . \qquad (3)$$

Here the 3-qubits are denoted by $A, B, C$, $\rho$ is the density matrix, $I$ denotes the unit $2\times 2$ matrix, $Tr_A$ represents the trace over qubit $A$ and $\otimes$ denotes the outer product. The density matrix of the 3 qubits system with MDS [1] is given as:

$$8\rho = (I)_A \otimes (I)_B \otimes (I)_C + \sum_{a,b,c=1}^{3} R_{a,b,c} (\sigma_a)_A \otimes (\sigma_b)_B \otimes (\sigma_c)_C$$
$$R_{a,b,c} = Tr\left(\rho (\sigma_a)_A \otimes (\sigma_b)_B \otimes (\sigma_c)_C\right) \qquad , \qquad (4)$$

$(\sigma_a)_A$ ($a = 1,2,3$), are the 3-components of the Pauli matrices for qubit $A$ and similarly for qubits $B$ and $C$. In the general 3-qubits MDS system 27 parameters are included in $R_{a,b,c}$. Eq. (4) represents a special case of the Hilbert-Schmidt (HS) decomposition of 3-qubits density matrices. Choosing different bases for the Pauli matrices of $A, B, C$, i. e, applying orthogonal transformations to $(\sigma_a)_A, (\sigma_b)_B, (\sigma_c)_C$ will turn $R_{a,b,c}$ into, say, $S_{p,q,r}$. It has been shown in our previous work [1] that if the sum of the absolute values of the HS parameters of the general 3-qubits state is not larger than 1, we can conclude that the density matrix is fully separable and we have an explicit expression for such separability. For the special case of 3-qubits MDS density matrices such sufficient condition for separability is given by [1]:

$$\sum_{a,b,c=1}^{3} |R_{a,b,c}| \leq 1 \quad , \qquad (5)$$

where $\sum_{a,b,c=1}^{3} |R_{a,b,c}|$ is considered as the $l_1$ norm of the tensor $R_{a,b,c}$ [6]. The crucial point here is that the form $\left(\sum_{a,b,c=1}^{3} |R_{a,b,c}|\right)$ is not invariant under orthogonal transformations and we can improve the condition for separabilty by using orthogonal transformations which will reduce this "separability form".

Compared to the separability problem of two qubits the separability problem for 3 qubits MDS density matrix becomes very complicated as tensors cannot be diagonalized. In order to overcome the non-diagonalization problem of tensors, methods of "unfolding of tensors into matrices" have been described in the literature [7-11]. These unfolding methods help us in the analysis of 3-qubits MDS systems, but as we are interested in application of such methods to density matrices (which was not the concern in [7-11]) special unfolding methods for this purpose are developed in the present work. While conditions for separability of 3-qubits have been treated by various other authors (see e.g. [12-22]) they have not applied the unfolding methods. Also the other authors have not given <u>explicitly separable forms</u> for the density matrices, which is done in the present work.

We analyze full separability properties of the 3-qubits MDS density matrices given by Eq. (4) by using unfolding of the tensor $R_{a,b,c}$. One way of unfolding the 3-qubits tensor $R_{a,b,c}(a,b,c=1,2,3)$ relative to $A$ means that you keep the parameter $a$ fixed as 1 or 2 or 3, (with the corresponding Pauli matrices $(\sigma_1)_A$, or $(\sigma_2)_A$, or $(\sigma_3)_A$) and then the parameters $R_{1,b,c}$, or $R_{2,b,c}$, or $R_{3,b,c}$ are considered, respectively, as matrices with $3\times 3$ dimension. Then, by using the singular-value-decomposition (SVD) [7, 8] for the matrices: $R_{1,b,c}$, $R_{2,b,c}$, and $R_{3,b,c}$ we get for each of them 3 singular values ($SV's$). We show that if the sum of these 9 SV's is not larger than 1, we can conclude that the density matrix is fully separable and we have an explicit separable form for the density matrix. Similar unfolding method can be made, relative to $B$ or $C$.

In another method we develop fully separable forms for 3 qubits MDS density matrices which are related to Frobenius ($l_2$) norms [6] of 9 triads of HS parameters. We show that if the sum of $l_2$ norms of the 9 HS triads is not larger than 1 then the 3-qubis MDS density matrix is fully separable, and we have another explicitly separable form for it.

For 3-qubits the density matrix may not be fully separable but may be biseparable, i.e., not genuinely entangled [23]. A condition for biseparability of 3-qubits MDS density matrices is obtained in the present work by the use of one qubit density matrix multiplied by Bell entangled states [2, 24-25] of the other two qubits. By using this method the 27 HS parameters are divided into 9 triads which are different from those used for full separability. If the sum of the nine $l_2$ norms for these different triads is not larger than 1, then we conclude that the density matrix is (at least) biseparable. We demonstrate the use of our methods in two examples comparing the conditions for full separability with the conditions for biseparability.

We apply the method of high order singular value decomposition (HOSVD) [7-11] for treatment of sufficient conditions for separability of 3-qubits MDS states. We demonstrate the application of this method in two examples.

For the general (non-MDS) 3-qubits density matrices, the HS decomposition includes 64 parameters which in a 4 dimensional notation may be written as: $R_{\mu,\nu,\kappa}$; $\mu,\nu,\kappa = 0,1,2,3$. Such terms include products of Pauli matrices $\sigma_i \, (i=1,2,3)$, and the unit operators $\sigma_0 \equiv I$. In various actual cases some of these parameters vanish. A sufficient condition for full separability is given by [1]:

$$\left( \sum_{\substack{\mu,\nu,\kappa=0 \\ \mu,\nu,\kappa \neq 0,0,0}}^{3} |R_{\mu,\nu,\kappa}| \right) \leq 1 \quad ; \quad R_{0,0,0} = 1 \quad , \tag{6}$$

but this condition may be improved. We demonstrate improvements in the condition for full separability by analyzing the system of GHZ state mixed with white noise and of W state mixed with white noise and get better conditions for full separability.

The present paper is arranged as follows: In Section II we describe fundamental properties of 3-qubits MDS density matrices. In Section III we describe unfolding methods for the 3-qubits MDS density matrices by which the tensor $R_{a,b,c}(a,b,c=1,2,3)$ is transformed into matrices [7-11]. Such unfolding processes are very useful in the numerical calculations related to the condition of separability of the density matrix (4). Explicit fully separable forms for the 3-qubits MDS density matrices, based on the $l_1$ norm are presented in Section IV. In Section V we describe fully separable forms for the 3-qubits MDS density matrices related to the $l_2$ norm. In this analysis the condition for full separability is given by the sum of the $l_2$ norms of 9 triads of HS parameters (where the 9 triads include the 27 $R_{a,b,c}(a,b,c=1,2,3)$ parameters). We make the calculations for full separability by this method, for the two examples treated in Section IV by the $l_1$ norm, and compare the results obtained by the two methods. In the same section we treat biseparability, where the 27 HS parameters are divided into 9 triads which are different from those used for full separability. If the sum of the nine $l_2$ norms for these different triads is not larger than 1 then we can conclude that the density matrix is at least biseparable and not genuinely entangled. In Section VI we develop the relations between 3 qubits MDS density matrices and the HOSVD method showing improvement of sufficient conditions for separability by using this method. For the general 3-qubits density matrices (non-MDS) which include outer products of $\sigma's$ with unit matrices, a sufficient condition for full separability is given by (6) but we can improve this condition. We demonstrate such improvements in section VII by analyzing the condition for full separability of GHZ state mixed with white noise and of W state mixed with white noise and we get better conditions for their full separability. In Section VIII we summarize our results and conclusions.

## II. Fundamental properties of 3-qubits MDS density matrices

For treating the 3-qubits MDS density matrix it is interesting, first, to show that the Peres-Horodecki (PH) criterion [26, 27] does not give information on such systems as the density matrix $\rho$ and its partial transpose (PT), $\rho(PT)$, have the same eigenvalues. For showing this result we write the 3-qubits MDS density matrix as

$$8\rho = (I)_A \otimes (I)_B \otimes (I)_C + R \quad . \tag{7}$$

Here R, given in a short notation, includes all the terms in the summation of Eq. (4). By performing the (full) transpose of $\rho$ into $\rho^T$, every $\sigma_y$ in Eq. (4) is transformed to $-\sigma_y$. This transformation does not change the eigenvalues ($\rho$ and $\rho^T$ have the same eigenvalues). By a $180^0$ unitary rotation of all qubits around the $y$ axis the eigenvalues of the density matrix are not changed, but $\sigma_x \to -\sigma_x, \sigma_z \to -\sigma_z$. We denote the resulting density matrix by $\rho^{TU}$. Here the superscript $TU$ represents transpose of (4), the whole density matrix, plus a unitary transformation. We emphasize that $\rho^{TU}$ and $\rho$ have the same eigenvalues. However, since we assumed an odd number of $\sigma$ we get $R \to -R$ by the TU transformation. Hence

$$8\rho^{TU} = (I) \otimes I \otimes I - R \quad . \tag{8}$$

On the other hand, the partial transpose plus a $180^0$ rotation around $y$ for one qubit (say qubit A) also yields

$$8\rho(PTU;A) = (I) \otimes I \otimes I - R \quad . \tag{9}$$

We find therefore that $\rho(PTU;A)$ has the same eigenvalues as $\rho$ so that the PH criterion [26, 27] does not give information for 3-qubits MDS states. This proof can easily be generalized for any odd number of qubits with MDS, where the eigenvalues of $\rho$ are equal to the eigenvalues of its PT transformation so that for such systems the PH criterion does not give information about entanglement.

A further conclusion comes from the fact (using the same argument) that for MDS density matrices with qubits of odd $n$ the eigenvalues of $(I)^{(n)} + R$ are the same as those of $(I)^{(n)} - R$; it follows that the eigenvalues of $\rho$ can be written as

$$Eigenvalues(\rho) = \frac{1 \pm r_i}{2^n} \quad (|r_i| \leq 1) \quad . \tag{10}$$

The eigenvalues of $R$ come in pairs $\pm r_i$ for any MDS density matrix which is of order odd n (including the 3-qubits as a special case for n=3) and are bounded by $\frac{1}{2^{n-1}}$.

The separability problem can also be related to Frobenius ($l_2$) norms which are given by the square root of sums of squared HS parameters [7, 8]. Let us prove the following relation for a 3 qubits MDS density matrix

$$\sum_{a,b,c=1}^{3} R_{a,b,c}^2 \leq 1 \quad . \tag{11}$$

We note first that

$$Tr\left[(8\rho)^2\right] = 8 + 8 \sum_{a,b,c=1}^{3} R_{a,b,c}^{2} \quad . \tag{12}$$

On the other hand $Tr\left[(8\rho)^2\right] = 64 \sum_{i=1}^{8} \lambda_i^2$ (13)

Here, $\lambda_i$ are the 8 eigenvalues of $\rho$. Since $0 \leq \lambda_i \leq 1/4$ (recalling that the 8 $r_i$ come in 4 pairs $\pm |r_i|$, as given by Eq. (10)) we write:

$$\lambda_i = \frac{1}{8} + q_i \quad ; \quad q_i = \frac{r_i}{8} \quad ; \quad |q_i| \leq \frac{1}{8} \quad ; \quad \sum_{i=1}^{8} q_i = 0 \quad ; \quad . \tag{14}$$

Hence

$$\sum_{i=1}^{8} \lambda_i^2 = \sum_{i=1}^{8}\left(\frac{1}{8} + q_i\right)^2 = \sum_{i=1}^{8}\left(\frac{1}{64}\right) + \sum_{i=1}^{8}\left(q_i^2\right) \leq \frac{1}{8} + \frac{1}{8} = \frac{1}{4} \quad . \tag{15}$$

By using Eqs. (12-15), we get Eq. (11). Eq. (11) may be generalized to any MDS density matrix with odd-n. Note that the equality in Eq. (11) holds only if

$$\lambda_i = \frac{1}{4} \ (i=1,2,3,4) \quad ; \quad \lambda_j = 0 \ (j=5,6,7,8) \quad . \tag{16}$$

According to Eq.(11), a necessary condition for (4) to be a density matrix is that the Frobenius norm of the sum of the 27 parameters, represented by the left side of Eq. (11), should not be larger than 1.

### III. Unfolding of the 3-qubits MDS tensor $R_{a,b,c}$ into matrices

In this section we describe unfolding processes by which tensors are unfolded into matrices. Such processes have been described in the literature [7-11] but the use of such unfolding processes becomes different in the present paper as we relate the analysis to density matrices, which was not the concern of the other works [7-11]

A certain unfolding of a tensor $R_{a,b,c}$ with dimension $n_1 \times n_2 \times n_3$ is obtained by assembling the $R's$ entries into a matrix with dimension $N_1 \times N_2 = n_1 \times n_2 \times n_3$ [7-11]. For our case $n_1 = n_2 = n_3 = 3$ and $N_1 = 3$, $N_2 = 9$. We denote the unfolded matrix of $R_{a,b,c}$ relative to qubit A as $R_{(b,c)}^{(a)} \equiv R_{(1)}$, and the matrix $R_{(1)}$ is arranged so that $R_{1,b,c}$ are inserted in the first row, $R_{2,b,c}$ are inserted in the second row and $R_{3,b,c}$ in the third row. The indices $(b,c) = 1,1; 1,2; 1,3; 2,1; 2,2; 2,3; 3,1; 3,2; 3,3$ are inserted into the $1, 2, \cdots, 9$ columns, respectively. By using this unfolding process $R_{a,b,c}$ is unfolded into:

$$R_{(1)} = R_{(b,c)}^{(a)} = \begin{Bmatrix} R_{111}, R_{112}, R_{113}, R_{121}, R_{122}, R_{123}, R_{131}, R_{132}, R_{133} \\ R_{211}, R_{212}, R_{213}, R_{221}, R_{222}, R_{223}, R_{231}, R_{232}, R_{233} \\ R_{311}, R_{312}, R_{313}, R_{321}, R_{322}, R_{323}, R_{331}, R_{332}, R_{333} \end{Bmatrix} . \quad (17)$$

One should note that $R_{(b,c)}^{(a)}$ is a matrix with $3 \times 9$ dimension while the matrices $R_{b,c}^{(a)}$ (without the brackets around $(b,c)$) related to the discussions about Pauli matrices in the next section, represent matrices with $3 \times 3$ dimension. The elements of the $3 \times 3$ matrix $R_{b,c}^{(a)}$ are composed of the $a'th$ row elements of $R_{(b,c)}^{(a)}$.

The tensor $R_{a,b,c}$ can be unfolded relative to qubit B by exchanging $R_{(b,c)}^{(a)}$ into $R_{(a,c)}^{(b)}$ so that for $(b) = 1, 2, 3$ the entries $R_{a,b,c}$ are inserted in the first, second and third row, respectively, and the entries related to $(a,c) = 1,1; 1,2; 1,3; 2,1; 2,2; 2,3; 3,1; 3,2; 3,3$ are inserted into the $1, 2, \cdots, 9$ columns, respectively. By using this unfolding $R_{(a,c)}^{(b)}$ is given as

$$R_{(2)} = R_{(a,c)}^{(b)} = \begin{Bmatrix} R_{111}, R_{112}, R_{113}, R_{211}, R_{212}, R_{213}, R_{311}, R_{312}, R_{313} \\ R_{121}, R_{122}, R_{123}, R_{221}, R_{222}, R_{223}, R_{321}, R_{322}, R_{323} \\ R_{131}, R_{132}, R_{133}, R_{231}, R_{232}, R_{233}, R_{331}, R_{332}, R_{333} \end{Bmatrix} , \quad (18)$$

Here again one should take into account that the matrix $R_{a,c}^{(b)}$ (without brackets around $a,c$) used in the nest section is of $3 \times 3$ dimension where its elements are composed from the $b'th$ row elements of the $3 \times 9$ matrix $R_{(a,c)}^{(b)}$. In a similar way the tensor $R_{a,b,c}$ can be unfolded relative to $c$ exchanging $R_{(b,c)}^{(a)}$, into $R_{(a,b)}^{(c)}$. For $(c) = 1, 2, 3$ the entries $R_{a,b,c}$ are inserted in the first, second and third row, respectively and the entries related to $(a,b) = 1,1; 1,2; 1,3; 2,1; 2,2; 2,3; 3,1; 3,2; 3,3$ are inserted into the $1, 2, \cdots, 9$ columns so that the unfolding to the matrix $R_{(a,b)}^{(c)}$ can be written explicitly as

$$R_{(3)} = R_{(a,b)}^{(c)} = \begin{Bmatrix} R_{111}, R_{121}, R_{131}, R_{211}, R_{221}, R_{231}, R_{311}, R_{321}, R_{331} \\ R_{112}, R_{122}, R_{132}, R_{212}, R_{222}, R_{232}, R_{312}, R_{322}, R_{332} \\ R_{113}, R_{123}, R_{133}, R_{213}, R_{223}, R_{233}, R_{313}, R_{323}, R_{333} \end{Bmatrix} . \tag{19}$$

Here again one should take into account that the matrix $R_{a,c}^{(c)}$ is of $3 \times 3$ dimension where its elements are obtained from the $c'th$ row of the matrix $R_{(a,b)}^{(c)}$.

For a general example of 3-qubits MDS state the 27 parameters $R_{a,b,c}$ can be inserted either in (17), or (18) or (19), giving $R_{(1)}$, or $R_{(2)}$, or $R_{(3)}$ matrix, respectively.

## IV. Explicitly separable forms for 3-qubits MDS density matrices related to reduction of the $l_1$ norm

A fully separable form for the density matrix (4) related to the $l_1$ norm can be given as:

$$8\rho_{ABC} =$$

$$(1/4) \sum_{a,b,c=1}^{3} |R_{a,b,c}| \cdot \begin{bmatrix} [\{(I)_A + (\sigma_a)_A\} \otimes \{(I)_B - (\sigma_b)_B\} \otimes \{(I)_C - sign(R_{abc})(\sigma_c)_C\}] + \\ [\{(I)_A + (\sigma_a)_A\} \otimes \{(I)_B + (\sigma_b)_B\} \otimes \{(I)_C + sign(R_{abc})(\sigma_c)_C\}] + \\ [\{(I)_A - (\sigma_a)_A\} \otimes \{(I)_B - (\sigma_b)_B\} \otimes \{(I)_C + sign(R_{abc})(\sigma_c)_C\}] + \\ [\{(I)_A - (\sigma_\alpha)_A\} \otimes \{(I)_B + (\sigma_\beta)_B\} \otimes \{(I)_C - sign(R_{abc})(\sigma_c)_C\}] \end{bmatrix}, \tag{20}$$

$$+ \left(1 - \sum_{a,b,c=1}^{3} |R_{a,b,c}|\right) \{(I)_A\} \otimes \{(I)_B\} \otimes \{(I)_c\}$$

Each expression in the curly brackets of (20) represents a pure state density matrix multiplied by 2. We get according to (20) that a sufficient condition for full separability is given by

$$\left(\sum_{a,b,c=1}^{3} |R_{a,b,c}|\right) \leq 1 . \tag{21}$$

This seems the simplest sufficient condition for full separability but it is not necessary and may be improved as follows.

The 3-qubits products of Eq. (4) can be written (say relative to qubit $A$) as

$$\sum_{a,b,c=1}^{3} R_{a,b,c} (\sigma_a)_A \otimes (\sigma_b)_B \otimes (\sigma_c)_C = \sum_{a=1}^{3} (\sigma_a)_A \otimes \sum_{b,c=1}^{3} R^{(a)}{}_{b,c} (\sigma_b)_B \otimes (\sigma_c)_C \ ; \ R^{(a)}{}_{b,c} \equiv R_{a,b,c} . \quad (22)$$

Notice that here $R^{(a)}{}_{b,c}$ ($a=1,2,3$) denotes a $3\times 3$ matrix. We note that the matrices $R^{(a)}{}_{b,c}$ are used here for the purpose of analyzing separability properties of the density matrix (4). We use now transformations which reduce the 27 HS parameters, $R_{a,b,c}$, to 9 parameters with smaller $l_1$ norm. Performing the SVD [7, 8] on the matrices $R^{(a)}{}_{b,c}$ in (22) we get:

$$\sum_{a=1}^{3} (\sigma_a)_A \otimes \sum_{b,c=1}^{3} R^{(a)}{}_{b,c} (\sigma_b)_B \otimes (\sigma_c)_C = \sum_{a=1}^{3} (\sigma_a)_A \otimes \sum_{i=1}^{3} R^{(a)}{}_i (\sigma_i)^{(a)}{}_B \otimes (\sigma_i)^{(a)}{}_C . \quad (23)$$

Here we used the SVD relation

$$\sum_{b,c} U^{(a)}{}_{b,i}{}^T R^{(a)}{}_{b,c} V^{(a)}{}_{j,c} = \delta_{i,j} R^{(a)}{}_i \quad , \quad (24)$$

where $R^{(a)}{}_i$ are the SV of $R^{(a)}{}_{b,c}$, $U^{(a)}$ and $V^{(a)}$ are $3\times 3$ real orthogonal matrices, and $(\sigma_i)^{(a)}{}_B = \sum_{b=1}^{3} U_{b,i} (\sigma_b)_B$, etc. . Taking absolute values in (24) we get

$$\left| R^{(a)}{}_i \right| = \left| \sum_{b,c} U^{(a)}{}_{i,b} V^{(a)}{}_{c,i} R^{(a)}{}_{b,c} \right| \leq \sum_{b,c} \left| U^{(a)}{}_{i,b} \right| \left| V^{(a)}{}_{c,i} \right| \left| R^{(a)}{}_{b,c} \right| \quad (25)$$

Performing the summation over $i$ we get

$$\sum_{i=1}^{3} \left| R^{(a)}{}_i \right| \leq \sum_{b,c=1}^{3} \sum_{i=1}^{3} \left| U^{(a)}{}_{i,b} \right| \left| V^{(a)}{}_{c,i} \right| \left| R^{(a)}{}_{b,c} \right| \quad . \quad (26)$$

$\left| U^{(a)}{}_{i,b} \right|$, for a certain $b$, and $\left| V^{(a)}{}_{c,i} \right|$, for a certain $c$, are unit vectors so that we get

$$\sum_{i=1}^{3} \left| U^{(a)}{}_{i,b} \right| \left| V^{(a)}{}_{c,i} \right| \leq 1 \quad . \quad (27)$$

Substituting (27) into (26) we get the relation

$$\sum_{i=1}^{3} \left| R^{(a)}{}_i \right| \leq \sum_{b,c=1}^{3} \left| R^{(a)}{}_{b,c} \right| \quad . \quad (28)$$

We find that the sum of the SV's absolute values $\sum_{i=1}^{3}\left|R^{(a)}_{i}\right|$ is smaller or equal to the $l_1$ norm of the matrix $R^{(a)}_{b,c}$. Since usually Eq. (27) is a strict inequality we expect a corresponding improvement in the sufficient condition.

Using the right hand side of Eq. (23) in Eq. (20) and using the general criterion (21) we find that under the condition (relative to A),

$$\sum_{i,a=1}^{3}\left|R^{(a)}_{i}\right|\leq 1, \qquad (29)$$

an explicit fully separable form for the 3-qubits MDS density matrix is obtained. In a similar way by using this procedure relative to B or C one gets, respectively,

$$\sum_{i,b=1}^{3}\left|R^{(b)}_{i}\right|\leq 1 \quad ; \quad \sum_{i,c=1}^{3}\left|R^{(c)}_{i}\right|\leq 1. \qquad (30)$$

One can choose the optimal condition for explicit full separability from the three conditions given by (29) and (30).

We demonstrate the present method, for improving the condition for separability by decreasing the $l_1$ norm, in the following two examples:

**Example 1:**

We have chosen, at random, 27 $R_{a,b,c}$ parameters given by:

$$\begin{aligned}
&R_{1,1,1}=0.0607\,;\,R_{1,1,2}=0.012\,;\,R_{1,1,3}=-0.0369\,;\,R_{1,2,1}=0.0216\,;\,R_{1,2,2}=0.0697\,;\\
&R_{1,2,3}=0.0952\,;\,R_{1,3,1}=0.0912\,;\,R_{1,3,2}=-0.0323\,;\,R_{1,3,3}=0.0344,\\
&R_{2,1,1}=0.0892\,;\,R_{2,1,2}=0.0489\,;\,R_{2,1,3}=0.0643\,;\,R_{2,2,1}=0.0377\,;\,R_{2,2,2}=-0.0451\,;\\
&R_{2,2,3}=0.0433\,;\,R_{2,3,1}=-0.0632\,;\,R_{2,3,2}=0.0381\,;\,R_{2,3,3}=-0.0675,\\
&R_{3,1,1}=0.0415\,;\,R_{3,1,2}=0.0305\,;\,R_{3,1,3}=0.0438\,;\,R_{3,2,1}=0.0425\,;\,R_{3,2,2}=0.0322\,;\\
&R_{3,2,3}=0.0671\,;\,R_{3,3,1}=0.0283\,;\,R_{3,3,2}=0.0514\,;\,R_{3,3,3}=-0.0673.
\end{aligned} \qquad (31)$$

By substituting these values in the density matrix (4) and performing all multiplications of Pauli matrices we arrive at the density matrix which has the eigenvalues:

$$8\lambda_1 = 0.553179\,;\,8\lambda_2 = 1.446821\,;\,8\lambda_3 = 1.353291\,;\,8\lambda_4 = 0.646709\,;$$
$$8\lambda_5 = 1.121965\,;\,8\lambda_6 = 0.878035\,;\,8\lambda_7 = 1.062880\,;\,8\lambda_8 = 0.937120 \quad . \tag{32}$$

We get 4 pairs of eigenvalues, with the relations:

$$\lambda_1 + \lambda_2 = \lambda_3 + \lambda_4 = \lambda_5 + \lambda_6 = \lambda_7 + \lambda_8 = 1/4 \quad . \tag{33}$$

This result is in agreement with Eq. (10).

The 27 parameters $R_{a,b,c}$ are inserted in the unfolded matrix $R_{(1)} = R_{(b,c)}^{(a)}$ given by (17) as:

$$R_{(1)} = \begin{Bmatrix} 0.0607, 0.012, -0.0369, 0.0216, 0.0697, 0.0952, 0.0912, -0.0323, 0.0344, \\ 0.0892, 0.0489, 0.0643, 0.0377, -0.0451, 0.0433, -0.0632, 0.0381, -0.0675, \\ 0.0415, 0.0305, 0.0438, 0.0425, 0.0322, 0.0671, 0.0283, 0.0514, -0,0673 \end{Bmatrix}. \tag{34}$$

Using (21) as a sufficient condition for separability, we get for the $R_{a,b,c}$ parameters of (31):

$$\left( \sum_{a,b,c=1}^{3} \left| R_{a,b,c} \right| \right) = 1.356 > 1 \quad . \tag{35}$$

This simple criterion fails to show full separability, but by using the SVD for the matrices $R^{(a)}_{b,c}$ as developed in Eqs. (22-29) we can use the relation: $\sum_{i,a=1}^{3} \left| R^{(a)}_i \right| \leq 1$ as a sufficient condition for full separability. For this purpose we calculate the $SV's$, $R^{(a)}_i$ ($a,i = 1,2,3$) of the matrices: $R^{(a)}_{b,c}$.

$$R_i^{(a=1)}(i=1,2,3) = SV \begin{pmatrix} 0.0607 & 0.012 & -0.0369 \\ 0.0216 & 0.0697 & 0.0952 \\ 0.0912 & -0.0323 & 0.0344 \end{pmatrix} = (0.126784, 0.107171, 0.050617)$$

$$R_i^{(a=2)}(i=1,2,3) = SV \begin{pmatrix} 0.0892 & 0.0489 & 0.0643 \\ 0.0377 & -0.0451 & 0.0433 \\ -0.0632 & 0.0381 & -0.0675 \end{pmatrix} = (0.154003, 0.078063, 0.002004) \quad . \tag{36}$$

$$R_i^{(a=3)}(i=1,2,3) = SV \begin{pmatrix} 0.0415 & 0.0305 & 0.0438 \\ 0.0425 & 0.0322 & 0.0671 \\ 0.0283 & 0.0514 & -0.0673 \end{pmatrix} = (0.110864, 0.087102, 0.003328)$$

The density matrix (4), with the 27 $R_{a,b,c}$ parameters given by (31), is fully separable as

$$\sum_{i,a=1}^{3} |R^{(a)}_{i}| = 0.7199 < 1 \qquad . \qquad (37)$$

Here the value 0.7199 has been obtained by the sum of the 9 SV's of Eq. (36).

While in the above analysis the 27 parameters $R_{a,b,c}$ were related to the 3 matrices $R_{b,c}^{(a)}$ ($a = 1,2,3$) similar analysis can be made if they will be related to $R_{a,c}^{(b)}$, or $R_{a,b}^{(c)}$. One can choose the optimal condition for explicit full separability from these 3 possibilities.

**Example 2**

We have chosen, at random, 9 $R_{a,b,c}$ parameters given by:

$$\begin{aligned} &R_{1,1,2} = 0.05\,;\, R_{1,1,3} = 0.22\,;\, R_{1,3,2} = 0.12\,;\, R_{1,3,3} = 0.2\,;\, R_{2,1,1} = 0.12\,;\\ &R_{2,2,1} = 0.3\,;\, R_{3,1,1} = 0.15\,;\, R_{3,2,2} = 0.25\,;\, R_{3,3,3} = 0.1 \end{aligned} \qquad , \qquad (38)$$

assuming that all other parameters $R_{a,b,c}$ vanish. Using these values in the density matrix (4) the calculated eigenvalues are given by:

$$\begin{aligned} &8\lambda_1 = 1.639266\,;\, 8\lambda_2 = 1.515047\,;\, 8\lambda_3 = 1.375857\,;\, 8\lambda_3 = 1.216394\,;\\ &8\lambda_5 = 0.360734\,;\, 8\lambda_6 = 0.484953\,;\, 8\lambda_7 = 0.624143\,;\, 8\lambda_8 = 0.783606 \end{aligned} \qquad . \qquad (39)$$

Again we note that the eigenvalues appear in pairs where

$$\lambda_i + \lambda_{i+4} = 1/4 \quad (i = 1,2,3,4) \quad \text{(in agreement with Eq. (10))}. \qquad (40)$$

Using (21) as a sufficient condition for separability we get for the $R_{a,b,c}$ parameters of (35):

$$\left( \sum_{a,b,c=1}^{3} |R_{a,b,c}| \right) = 1.51 > 1 \qquad . \qquad . \qquad (41)$$

By using the $SV's$ for the matrices $R^{(a)}_{b,c}$, we can try to improve the condition for separability:

$$R_i^{(a=1)}(i=1,2,3) = SV \begin{pmatrix} 0 & 0.05 & 0.22 \\ 0 & 0 & 0 \\ 0 & 0.12 & 0.2 \end{pmatrix} = (0.32044, 0.05118, 0)$$

$$R_i^{(a=2)}(i=1,2,3) = SV \begin{pmatrix} 0.12 & 0 & 0 \\ 0.3 & 0 & 0 \\ 0 & 0 & 0 \end{pmatrix} = (0.32311, 0, 0) \qquad . \qquad (42)$$

$$R_i^{(a=3)}(i=1,2,3) = SV \begin{pmatrix} 0.15 & 0 & 0 \\ 0 & 0.25 & 0 \\ 0 & 0 & 0.1 \end{pmatrix} = (0.15, 0.25, 0.1)$$

By adding the 9 SV's of Eq. (42) we check the condition (29) for full separability and get:

$$\sum_{i,a=1}^{3} |R^{(a)}_i| = 1.19473 > 1 \ . \qquad (43)$$

So the sufficient condition for separability is still not satisfied but the $l_1$ norm obtained in (43) is much smaller than that obtained in (41). By making the analysis of this example for $R^{(b)}_{a,c}$ or $R^{(c)}_{a,b}$ we find also for such cases that the condition for full separability is not satisfied.

## V. Full separability and biseparability for 3-qubits MDS density matrices related to the $l_2$ norm

A fully separable-like form for the density matrix (4) related to the $l_2$ norms can be given as:

$$8\rho_{ABC} = \frac{1}{4} \left[ \sum_{a,b=1}^{3} \sqrt{R_{a,b,1}^2 + R_{a,b,2}^2 + R_{a,b,3}^2} \begin{bmatrix} \begin{Bmatrix} [\{(I)_A + (\sigma_a)_A\} \otimes \{(I)_B + (\sigma_b)_B\}] \\ +[\{(I)_A - (\sigma_a)_A\} \otimes \{(I)_B - (\sigma_b)_B\}] \end{Bmatrix} \otimes \left\{ (I)_C + \frac{(R_{a,b,1}(\sigma_x) + R_{a,b,2}(\sigma_y) + R_{a,b,3}(\sigma_z))_C}{\sqrt{R_{a,b,1}^2 + R_{a,b,2}^2 + R_{a,b,3}^2}} \right\} \\ + \begin{Bmatrix} [\{(I)_A - (\sigma_a)_A\} \otimes \{(I)_B + (\sigma_b)_B\}] \\ +[\{(I)_A + (\sigma_\alpha)_A\} \otimes \{(I)_B - (\sigma_\beta)_B\}] \end{Bmatrix} \otimes \left\{ (I)_C - \frac{(R_{a,b,1}(\sigma_x) + R_{a,b,2}(\sigma_y) + R_{a,b,3}(\sigma_z))_C}{\sqrt{R_{a,b,1}^2 + R_{a,b,2}^2 + R_{a,b,3}^2}} \right\} \end{bmatrix} \right.$$

$$\left. + \left( 1 - \sum_{a,b=1}^{3} \left( \sum_{c=1}^{3} R_{a,b,c}^2 \right)^{1/2} \right) \{(I)_A\} \otimes \{(I)_B\} \otimes \{(I)_C\} \right]. \qquad (44)$$

Each term in the curly brackets of (44) represents a pure state density matrix multiplied by 2. It is straightforward to show that the complicated separable form of Eq. (44), is reduced to the density matrix (4) by manipulating all the cross products in this equation. According to Eq. (44) a sufficient condition for full separability is given by

$$\sum_{a,b=1}^{3}\left(\sum_{c=1}^{3} R_{a,b,c}^{2}\right)^{1/2} \leq 1 \quad . \tag{45}$$

In (45), $\left(\sum_{c=1}^{3} R_{a,b,c}^{2}\right)^{1/2}$ for certain $a,b$ values, represents the $l_2$ norm of a triad of HS parameters $(c=1,2,3)$. If the sum of $l_2$ norms over 9 triads of HS parameters (a, b=1, 2, 3) is not larger than 1 we can conclude that the density matrix (4) is fully separable and we have the explicit form (44) for its full separability. While the explicit form (20) of the density matrix (4) gives the sufficient condition for full separability $\left(\sum_{a,b,c=1}^{3}|R_{a,b,c}|\right) \leq 1$, related to the $l_1$ norm, the explicit form (44) of the density matrix (4) gives the sufficient condition for full separability related to the $l_2$ norm by (45) where the left side of this equation includes sum of 9 $l_2$ norms of triads. Note that Eq. (45) may hold when (21) is not satisfied.

Using the criterion (45) for separability we get for the above, <u>example 1</u>, represented by the matrix $R_{(1)}$ of (34):

$$\begin{aligned}&\left(0.0607^{2}+0.012^{2}+0.0369^{2}\right)^{1/2}+\left(0.0216^{2}+0.0697^{2}+0.0952^{2}\right)+\left(0.0912^{2}+0.0323^{2}+0.0344^{2}\right)+\\&(0.0892^{2}+0.0489^{2}+0.0643^{2})^{1/2}+\left(0.0377^{2}+0.0451^{2}+0.0433^{2}\right)^{1/2}+\left(0.0632+0.0381^{2}+0.0675^{2}\right)^{1/2}+\\&\left(0.0415^{2}+0.0305^{2}+0.0438^{2}\right)^{1/2}+\left(0.0425^{2}+0.0322^{2}+0.0671^{2}\right)^{1/2}+\left(0.0283^{2}+0.0514^{2}+0.0673^{2}\right)^{1/2}\\&=0.829456<1\end{aligned} \tag{46}$$

We find that the density matrix (2) in example 1 can be presented by the explicitly separable form of Eq. (44). For example 2 we find the relation $\sum_{a,b=1}^{3}\left(\sum_{c=1}^{3} R_{a,b,c}^{2}\right)^{1/2} > 1$ so that the density matrix for this example cannot be presented by the separable form (44). This does not mean that the state is not fully separable as the condition for full separability might be improved by other methods.

In our previous work [2] we treated <u>biseparability</u> of 3-qubits MDS density matrix by using one qubit density matrices multiplied by entangled Bell states of the other two qubits [2, 24, 25]. Let us show the biseparability obtained for the following simple 3-qubits MDS density:

$$8\rho_1 = (I)_A \otimes (I)_B \otimes (I)_C + \\ R_{111}(\sigma_x)_A \otimes (\sigma_x)_B \otimes (\sigma_x)_C + R_{222}(\sigma_y)_A \otimes (\sigma_y)_B \otimes (\sigma_y)_C + R_{333}(\sigma_z)_A \otimes (\sigma_z)_B \otimes (\sigma_z)_C \quad . \tag{47}$$

The 8 eigenvalues of this density matrix are given by

$$\lambda_1 = \lambda_2 = \lambda_3 = \lambda_4 = (1/8)\left[1 + \sqrt{\sum_{i=1}^{3} R_{iii}^2}\right] \quad ; \\ \lambda_5 = \lambda_6 = \lambda_7 = \lambda_8 = (1/8)\left[1 - \sqrt{\sum_{i=1}^{3} R_{iii}^2}\right] \quad . \tag{48}$$

Hence it is a density matrix when $R_{iii}$ are within the unit sphere, i.e. when

$$\sum_{i=1}^{3} R_{iii}^2 \leq 1 \quad . \tag{49}$$

The sufficient condition for full separability [1] is given by

$$\sum_{i=1}^{3} |R_{iii}| \leq 1 \quad . \tag{50}$$

Explicit biseparable expression for the density matrix (47) is given by:

$$8\rho_1 = \\ \left\{\begin{array}{l} \left[(I)_A - R_{111}(\sigma_x)_A + R_{222}(\sigma_y)_A + R_{333}(\sigma_z)_A\right] \otimes \left[|\Phi^{(-)}\rangle_{BC} \langle\Phi^{(-)}|_{BC}\right] + \\ \left[(I)_A + R_{111}(\sigma_x)_A - R_{222}(\sigma_y)_A + R_{333}(\sigma_z)_A\right] \otimes \left[|\Phi^{(+)}\rangle_{BC} \langle\Phi^{(+)}|_{BC}\right] + \\ \left[(I)_A + R_{111}(\sigma_x)_A + R_{222}(\sigma_y)_A - R_{333}(\sigma_z)_A\right] \otimes \left[|\Psi^{(+)}\rangle_{BC} \langle\Psi^{(+)}|_{BC}\right] + \\ \left[(I)_A - R_{111}(\sigma_x)_A - R_{222}(\sigma_y)_A - R_{333}(\sigma_z)_A\right] \otimes \left[|\Psi^{(-)}\rangle_{BC} \langle\psi^{(-)}|_{BC}\right] \end{array}\right\} \quad . \tag{51}$$

Here $|\Phi^{(-)}\rangle_{BC}$ $|\Phi^{(+)}\rangle_{BC}$, $|\Psi^{(+)}\rangle_{BC}$ and $|\Psi^{(-)}\rangle_{BC}$ are the Bell states [24, 25] of the qubits pair $B$ and $C$ expanded in terms of Pauli matrices as:

$$4|\Phi^{(-)}\rangle_{BC}\langle\Phi^{(-)}|_{BC} = \left[(I)_B \otimes (I)_C - (\sigma_x)_B \otimes (\sigma_x)_C + (\sigma_y)_B \otimes (\sigma_y)_C + (\sigma_z)_B \otimes (\sigma_z)_C\right] ;$$
$$4|\Phi^{(+)}\rangle_{BC}\langle\Phi^{(+)}|_{BC} = \left[(I)_B \otimes (I)_C + (\sigma_x)_B \otimes (\sigma_x)_C - (\sigma_y)_B \otimes (\sigma_y)_C + (\sigma_z)_B \otimes (\sigma_z)_C\right] ;$$
$$4|\Psi^{(+)}\rangle_{BC}\langle\Psi^{(+)}|_{BC} = \left[(I)_B \otimes (I)_C + (\sigma_x)_B \otimes (\sigma_x)_C + (\sigma_y)_B \otimes (\sigma_y)_C - (\sigma_z)_B \otimes (\sigma_z)_C\right] ;(52)$$
$$4|\Psi^{(-)}\rangle_{BC}\langle\Psi^{(-)}|_{BC} = \left[(I)_B \otimes (I)_C - (\sigma_x)_B \otimes (\sigma_x)_C - (\sigma_y)_B \otimes (\sigma_y)_C - (\sigma_z)_B \otimes (\sigma_z)_C\right] ;$$

Equations (51) and (52) represent a biseparable density matrix, under the condition

$$\sqrt{R_{111}^2 + R_{222}^2 + R_{333}^2} \leq 1 \quad , \tag{53}$$

which is equivalent to the condition (49) for $\rho$ of Eq. (47) to be a density matrix.

To treat biseparability of the general case of MDS density matrix given by (4) (up to 27 MDS terms), we can divide $\sum_{a,b,c=1}^{3} R_{a,b,c}(\sigma_a)_A \otimes (\sigma_b)_B \otimes (\sigma_c)_C$ into 9 groups of triads. Starting with Eq. (47) we can apply 8 transformations to the Pauli matrices of each qubit, obtaining 8 triads with corresponding HS parameters. Each triad may be treated as in Eqs. (51-52). Together with Eq. (47) they include the 27 $R_{l,m,n}$ parameters. The relevant triads are indicated in Eq. (54) below. Therefore the sufficient condition for biseparability of Eq. (4) becomes that the sum of the Frobenius norms of the 9 (at most) triads of MDS-parameters is not larger than 1. Such condition is sufficient for biseparability but the sufficient condition for biseparability may perhaps be improved by other methods.

The final conclusion from this analysis is that a sufficient condition for biseparability of a 3-qubits MDS density matrix is that the sum of $l_2$ norms over 9 triads of HS parameters, is not larger than 1, but the 9 HS triads are different from those used for full separability in Eqs. (44-45). A sufficient condition for biseparability of 3–qubits MDS density matrix in the general case of 27 HS parameters is given by

$$\sqrt{R_{111}^2 + R_{222}^2 + R_{333}^2} + \sqrt{R_{132}^2 + R_{321}^2 + R_{213}^2} + \sqrt{R_{123}^2 + R_{312}^2 + R_{231}^2}$$
$$+\sqrt{R_{112}^2 + R_{223}^2 + R_{331}^2} + \sqrt{R_{121}^2 + R_{232}^2 + R_{313}^2} + \sqrt{R_{133}^2 + R_{322}^2 + R_{211}^2} \qquad (54)$$
$$+\sqrt{R_{113}^2 + R_{221}^2 + R_{332}^2} + \sqrt{R_{212}^2 + R_{131}^2 + R_{323}^2} + \sqrt{R_{311}^2 + R_{233}^2 + R_{122}^2} \leq 1$$

Obviously conditions (45) and (54) are bounded from below by condition (11).

By substituting the 27 HS parameters of Eq. (31) (Example 1), in Eq. (54) we get the condition for biseparability:

$$0.7042 < 1 \quad , \qquad (55)$$

which is better than the conditions for full separability: $0.7199 < 1$ in (37) and $0.8295 < 1$ in (46). By assuming that the HS parameters of (31) are increased by a factor $1/0.71$, the sufficient conditions for full separability, given by Eqs. (29), and (45) are not satisfied but the sufficient condition for biseparability given by Eq. (54) is satisfied.

By substituting the 9 HS parameters of Eq. (38) (Example 2) in Eq. (55) we get for the left hand side of (54)

$$1.070 > 1 \quad , \qquad (56)$$

which is not sufficient for biseparability but is better than the condition for full separability $\sum_{i,a=1}^{3} |R^{(a)}_i| = 1.19473 > 1$ given in Eq. (43). By reducing the HS parameters, by a factor 0.9, the condition for biseparability is satisfied but not the condition for full separability.

## VI. Application of the high order singular value decomposition (HOSVD) for improving the sufficient condition for full separability of 3-qubits MDS density matrices

In section IV we have shown that by using special unfolding methods, the MDS tensors are converted into square matrices and by applying singular values decompositions (SVD) to these matrices the number of the HS parameters in the general case is reduced to 9 and under the condition that the sum of absolute values of these parameters (the $l_1$ norm) is not larger than 1, we conclude that the density matrix is fully separable and we get explicitly separable forms for the density matrices. It is interesting to see how this analysis can be extended by relating it to the method of high order singular value decomposition (HOSVD) [7-11].

In the HOSVD method we use the SVD for the matrices $R_{(1)}, R_{(2)}$ and $R_{(3)}$ (given, respectively, by (17), (18) and (19)):

$$R_{(1)} = U_1 \Sigma_1 V_1^T \quad ; \quad R_{(2)} = U_2 \Sigma_2 V_2^T \quad ; \quad R_{(3)} = U_3 \Sigma_3 V_3^T \quad . \tag{57}$$

Here, the subscripts 1, 2, 3 refer to the unfolded matrices: $R_{(1)} ; R_{(2)} ; R_{(3)}$ , respectively. The matrices $U_1, U_2, U_3$ are of $3\times 3$ dimension. The singular matrices $\Sigma_1, \Sigma_2, \Sigma_3$ are of $3\times 9$ dimension and the matrices $V_1, V_2, V_3$ are of $9\times 9$ dimension. The singular matrices $\Sigma_i$ are of the form

$$\Sigma_i = \begin{pmatrix} s_1(R_{(i)}) & 0 & 0 & 0 & 0 & 0 & 0 & 0 & 0 \\ 0 & s_2(R_{(i)}) & 0 & 0 & 0 & 0 & 0 & 0 & 0 \\ 0 & 0 & s_3(R_{(i)}) & 0 & 0 & 0 & 0 & 0 & 0 \end{pmatrix} \quad , \tag{58}$$

$s_1(R_i), s_2(R_i), s_3(R_i)$ are the three singular values of $R_{(i)}$ $(i=1,2,3)$. The tensor $R_{a,b,c}$ is related to the core tensor $S$ by the transformation [7-11],

$$R_{a,b,c} = \sum_{p=1}^{3} \sum_{q=1}^{3} \sum_{r=1}^{3} S_{p,q,r} U_1(a,p) U_2(b,q) U_3(c,r) \quad , \tag{59}$$

and by the inverse transformation

$$S_{p,q,r} = \sum_{a=1}^{3} \sum_{b=1}^{3} \sum_{n=c=1}^{3} R_{a,b,c} U_1(a,p)^T U_2(b,q)^T U_3(c,q)^T \quad . \tag{60}$$

While direct calculation of the core tensor $S$ (which includes the above Tucker products [7-11]) is quite complicated, for our purpose of calculating sufficient conditions for full separability, it is enough to calculate $S_{(1)}$ which is the unfolding of the core tensor $S$ relative to qubit $A$, and is given by [7-11]

$$S_{(1)} \equiv S_{(q,r)}^{(p)} = \Sigma_1 V_1 (U_3 \otimes U_2) \quad . \tag{61}$$

In a similar way we can calculate $S_{(2)}$ or $S_{(3)}$ which are the unfolding of the core tensor $S$ relative to qubit $B$, or $C$, respectively:

$$S_{(2)} \equiv S_{(p,r)}^{(q)} = \Sigma_2 V_2 (U_3 \otimes U_1) \quad ; \quad S_{(3)} \equiv S_{(p,q)}^{(r)} = \Sigma_3 V_3 (U_2 \otimes U_1) \quad . \quad (62)$$

The unfolded matrix $S_{(i)}$ has various special properties [7-11] including orthogonality between its rows. Also $\sqrt{\sum_{q,r=1}^{3} \left(S_{(q,r)}^{(p)}\right)^2}$ ($p=1,2,3$) are equal, respectively, to the singular values, $s_1(R_{(1)})$, $s_2(R_{(1)})$, $s_3(R_{(1)})$, of $\Sigma_1$. Similar properties hold relative to $S_{(2)}$, or $S_{(3)}$. Applying the orthogonal transformations, $U^T_{(1)}, U^T_{(2)}, U^T_{(3)}$ to $\vec{\sigma}_A, \vec{\sigma}_B, \vec{\sigma}_C$, respectively, means choosing new bases for the Pauli matrices: $\vec{\tilde{\sigma}}_A, \vec{\tilde{\sigma}}_B, \vec{\tilde{\sigma}}_C$. In terms of these, Eq. (4) becomes

$$8\rho = (I)_A \otimes (I)_B \otimes (I)_C + \sum_{p,q,r=1}^{3} S_{p,q,r} (\tilde{\sigma}_p)_A \otimes (\tilde{\sigma}_q)_B \otimes (\tilde{\sigma}_r)_C \quad . \quad (63)$$

All the various previous formulas may be written in terms of $S_{p,q,r}$ instead of $R_{a,b,c}$.

We demonstrate now the use of the HOSVD method, for improving the condition for full separability, in two examples:

**1) MDS density matrix with 27 equal HS parameters**

For this case $R_{a,b,c} = \alpha$ and the simplest condition for full separabilty (Eq. (5)) yields: $\alpha \leq 1/27$
The simplest condition for full separability according to the $l_2$ norm (Eq. (45)) yields: $\alpha \leq 1/9\sqrt{3}$.

This is also the condition for biseparability (Eq. (54))

The unfolding of the 3-qubits density matrix relative to qubit $A$ can be written as

$$8\rho = (I)_A \otimes (I)_B \otimes (I)_C + \sum_{a=1}^{3} (\sigma_a)_A \otimes \sum_{b,c=1}^{3} R^{(a)}_{b,c} (\sigma_b)_B \otimes (\sigma_c)_C \; ; R^{(a)}_{b,c} \equiv R_{a,b,c} = \alpha \quad . (64)$$

Then the 3 matrices $R^{(a)}_{b,c}$ are given by

$$R^{(a)}{}_{b,c} = \begin{pmatrix} \alpha & \alpha & \alpha \\ \alpha & \alpha & \alpha \\ \alpha & \alpha & \alpha \end{pmatrix} \quad ; \quad a = 1,2,3 \quad . \tag{65}$$

By calculating the SV's, of $R^{(a)}{}_{b,c}$, of Eq. (65), we get

$$R_1^{(a)} = 3\alpha \quad ; \quad R_2^{(a)} = R_3^{(a)} = 0 \quad ; \quad a = 1,2,3 \quad . \tag{66}$$

Then we have

$$\sum_{a=1}^{3}\sum_{i=1}^{3} R_i^{(a)} = 9\alpha \quad . \tag{67}$$

Therefore a sufficient condition for full separability is now given by

$$\alpha \le 1/9 \quad . \tag{68}$$

The eigenvalues of the density matrix $\rho$ in this example are given by:

$$\begin{aligned} \lambda_1 &= \lambda_2 = \lambda_3 = \lambda_4 = \frac{1+3\sqrt{3}\,\alpha}{8} \quad ; \\ \lambda_5 &= \lambda_6 = \lambda_7 = \lambda_8 = \frac{1-3\sqrt{3}\,\alpha}{8} \end{aligned} \quad . \tag{69}$$

We have a density matrix under the condition $3\sqrt{3}\,|\alpha| \le 1$. In the region: $1/9 < \alpha \le 1/3\sqrt{3}$ Eq. (68) for full separability does not hold. We will show now that this condition can be greatly improved by the use of HOSVD, so that in the whole region that we have a density matrix it is fully separable.

The unfolded matrices: $R_{(1)}$, $R_{(2)}$, $R_{(3)}$ of Eqs. (17-19) are given in the present example by

$$R_{(1)} = R_{(2)} = R_{(3)} = \begin{pmatrix} \alpha & \alpha & \alpha & \alpha & \alpha & \alpha & \alpha & \alpha & \alpha \\ \alpha & \alpha & \alpha & \alpha & \alpha & \alpha & \alpha & \alpha & \alpha \\ \alpha & \alpha & \alpha & \alpha & \alpha & \alpha & \alpha & \alpha & \alpha \end{pmatrix} \quad . \tag{70}$$

The high order transformed matrix $S_{(1)}$ is calculated by Eq. (61) where $\Sigma_1$ and $V_1$ are obtained by the SVD of $R_{(1)}$, $U_3$ and $U_2$ are calculated by the SVD of $R_{(3)}$ and $R_{(2)}$, respectively.

After straightforward calculations we get for this example

$$S_{(1)} = S_{(2)} = S_{(3)} = \begin{pmatrix} 3\sqrt{3}\alpha & 0 & 0 & 0 & 0 & 0 & 0 & 0 & 0 \\ 0 & 0 & 0 & 0 & 0 & 0 & 0 & 0 & 0 \\ 0 & 0 & 0 & 0 & 0 & 0 & 0 & 0 & 0 \end{pmatrix}. \qquad (71)$$

So, the condition for full separability is $3\sqrt{3}|\alpha| \leq 1$ which is equivalent to the condition for the present example to be a density matrix.

**2) Three-qubits MDS density matrices with 27 different HS parameters**

Let us assume that we have 3-qubits MDS density matrix with the following unfolding matrix $R_{(1)}$ of Eq. (17) relative to qubit $A$:

$$R_{(1)} = R_{(b,c)}^{(a)} = \begin{cases} 0.09105, 0.018, -0.05535, 0.0324, 0.10455, 0.1428, 0.1368, -0.04845, 0.0516 \\ 0.1338, 0.0735, 0.09645, 0.05655, -0.0675, 0.06495, -0.0948, 0.05715, -0.10125 \\ 0.06225, 0.04575, 0.0657, 0.06375, 0.0483, 0.10065, 0.04245, 0.0771, -0.10095 \end{cases}. \qquad (72)$$

Since we have chosen here the HS parameters to be 1.5 times the HS parameters of Eq. (31), we get here the eigenvalues

$$8\lambda_1 = 0.329769; \; 8\lambda_2 = 1.670231; 8\lambda_3 = 1.529937; \; 8\lambda_4 = 0.470063;$$
$$8\lambda_5 = 1.182948; 8\lambda_6 = 0.817052; 8\lambda_7 = 1.09432; 8\lambda_8 = 0.90568 \qquad (73)$$

By using the SVD for the matrices $R^{(a)}_{b,c}$, as developed in Eqs. (22-29) we can use here, again, the relation: $\sum_{i,a=1}^{3} |R^{(a)}_i| \leq 1$ as sufficient condition for full separability. Since the HS parameters in Eq. (72) are increased by a factor 1.5 relative to the HS parameters of Eq. (31) we get here the relation : $\sum_{i,a=1}^{3} |R^{(a)}_i| = 0.7199 \cdot 1.5 = 1.07985 > 1$. So the sufficient condition for separability

$\sum_{i,a=1}^{3} |R^{(a)}_i| \leq 1$ is not satisfied for $R_{(b,c)}^{(a)}$ of Eq. (72).

We would like to show that the sufficient condition for full separability is improved by using the HOSVD, following the relation $S_{(1)} = \Sigma_1 V_1 (U_3 \otimes U_2)$ by which we exchange $R_{(1)}$ into $S_{(1)}$. $\Sigma_1$, and $V_1$ are calculated by the SVD of the unfolded matrix $R_{(1)}$ while $U_2$ and $U_3$ are calculated by the SVD of the unfolded matrices $R_{(2)}$, and $R_{(3)}$, respectively. The 27 HS parameters are ordered in unfolded matrices according to Eqs. (17-19). A straightforward calculation gives

$$S_{(1)} = S_{(q,r)}^{(p)}$$
$$\begin{Bmatrix} 0.1333, -0.0959, 0.0272, 0.0612, 0.1577, -0.0969, 0.0121, 0.1435, -0.0974 \\ -0.0487, 0.0406, -0.1341, 0.0159, -0.1418, -0.0898, 0.0365, 0.1211, -0.0915 \\ 0.0293, -0.0031, -0.0372, -0.0133, 0.0140, 0.0142, 0.0812, -0.0018, 0.0284 \end{Bmatrix}. \quad (74)$$

The Frobenius norms of the 9 terms in the first, second and the third row in Eq. (74) are equal the singular values $s(1), s(2), s(3)$ of $R_{(1)}$ and these rows are orthogonal. One should notice that the $p'th$ row ($p = 1, 2, 3$) of Eq. (74) includes 9 terms where each 3 of them are inserted in the first, second and third row, of the matrix $S_{q,r}^{(p)}$, respectively. By calculating the SVD of these matrices, we get:

$$S_i^{(p=1)}(i=1,2,3) = SV \begin{pmatrix} 0.1333 & -0.0959 & 0.0272 \\ 0.0612 & 0.1577 & -0.0969 \\ 0.0121 & 0.1435 & -0.0974 \end{pmatrix} = (0.27113, 0.14931, 0.01152)$$

$$S_i^{(p=2)}(i=1,2,3) = SV \begin{pmatrix} -0.0487 & 0.0406 & -0.1341 \\ 0.0159 & -0.1418 & -0.0898 \\ 0.0365 & 0.1211 & -0.0915 \end{pmatrix} = (0.19841, 0.17771, 0.06205)$$

$$S_i^{(p=3)}(i=1,2,3) = SV \begin{pmatrix} 0.0293 & -0.0031 & -0.0372 \\ -0.0133 & 0.0140 & 0.0142 \\ 0.0812 & -0.0018 & 0.0284 \end{pmatrix} = (0.08855, 0.04746, 0.01167)$$

(75)

The sum of the 9 SV's gives

$$(0.27113, 0.14931, 0.01152) + (0.19841, 0.17771, 0.06205) +$$
$$(0.08855, 0.04746, 0.01167) = 0.43196 + 0.43817 + 0.1476 = 1.01773 \quad . \quad (76)$$

This sum is smaller than the sum $\sum_{i,a=1}^{3} |R^{(a)}_i| = 0.7199 \cdot 1.5 = 1.07985$. If we were to reduce the HS parameters of (72), say by factor 0.98, the condition for full separability will be satisfied, in terms of the $S$ parameters (Eq. (63)), but not in terms of the $R$ parameters.

## VII. Explicitly separable forms for GHZ state and W state mixed with white noise.

An arbitrary 3-qubits density matrix can be written as

$$8\rho_{ABC} = \sum_{\mu,\nu,\kappa=0}^{3} R_{\mu,\nu,\kappa}(\sigma_\mu) \otimes (\sigma_\nu) \otimes (\sigma_\kappa) \quad ; \quad R_{0,0,0} = 1 \quad . \tag{77}$$

The Hermiticity of $\rho$ is equivalent to the condition that the HS parameters $R_{\mu,\nu,\kappa}$ are real. In the general case Eq. (77) includes 63 HS parameters, but some of these parameters may vanish. The simplest condition for full separability of the density matrix (77) is given by [1]

$$\sum_{\substack{\mu,\nu,\kappa=0 \\ \mu,\nu,\kappa \neq 0,0,0}}^{3} |R_{\mu,\nu,\kappa}| \leq 1 \quad ; \quad R_{0,0,0} = 1 \quad . \tag{78}$$

But, usually this condition can be improved very much.

Let us treat, separately, the system of *GHZ* state mixed with white noise and that of *W* state mixed with white noise.

A GHZ state can be given as:

$$|\psi\rangle_{GHZ} = \frac{1}{\sqrt{2}}\left[|0\rangle_A \otimes |0\rangle_B \otimes |0\rangle_C + |1\rangle_A \otimes |1\rangle_B \otimes |1\rangle_C\right] \quad . \tag{79}$$

The HS decomposition of the density matrix $|\psi\rangle\langle\psi|_{GHZ}$ is given by [1,3]:

$$\begin{aligned}8\rho(GHZ) = &(I)_A \otimes (I)_B \otimes (I)_C + (\sigma_x)_A \otimes (\sigma_x)_B \otimes (\sigma_x)_C + (I)_A \otimes (\sigma_z)_B \otimes (\sigma_z)_C \\ &(\sigma_z)_A \otimes (I)_B \otimes (\sigma_z)_C + (\sigma_z)_A \otimes (\sigma_z)_B \otimes (I)_C - (\sigma_x)_A \otimes (\sigma_y)_B \otimes (\sigma_y)_C \\ &-(\sigma_y)_A \otimes (\sigma_x)_B \otimes (\sigma_y)_C - (\sigma_y)_A \otimes (\sigma_y)_B \otimes (\sigma_x)_C\end{aligned} \tag{80}$$

This density matrix with probability $p$ mixed with white noise is given by

$$\rho(GHZ; mixed) = p\rho(GHZ) + (1-p)(I)_A \otimes (I)_B \otimes (I)_C \quad . \tag{81}$$

Inserting Eq. (80) into Eq. (81), we get:

$$8\rho(GHZ;mixed) = p\begin{Bmatrix} (I)_A \otimes (I)_B \otimes (I)_C + (\sigma_x)_A \otimes (\sigma_x)_B \otimes (\sigma_x)_C + (I)_A \otimes (\sigma_z)_B \otimes (\sigma_z)_C \\ (\sigma_z)_A \otimes (I)_B \otimes (\sigma_z)_C + (\sigma_z)_A \otimes (\sigma_z)_B \otimes (I)_C - (\sigma_x)_A \otimes (\sigma_y)_B \otimes (\sigma_y)_C \\ -(\sigma_y)_A \otimes (\sigma_x)_B \otimes (\sigma_y)_C - (\sigma_y)_A \otimes (\sigma_y)_B \otimes (\sigma_x)_C \end{Bmatrix}$$
$$+(1-p)(I)_A \otimes (I)_B \otimes (I)_C \quad (82)$$

An explicitly separable form for $8\rho(GHZ;mixed)$ is given by:

$$8\rho(GHZ;mixed) =$$

$$\frac{p}{4} \cdot \begin{bmatrix} \begin{Bmatrix} (I+\sigma_x)_A \otimes (I-\sigma_x)_B \otimes (I-\sigma_x)_C + (I+\sigma_x)_A \otimes (I+\sigma_x)_B \otimes (I+\sigma_x)_C \\ +(I-\sigma_x)_A \otimes (I-\sigma_x)_B \otimes (I+\sigma_x)_C + (I-\sigma_x)_A \otimes (I+\sigma_x)_B \otimes (I-\sigma_x)_C \end{Bmatrix} \\ + \begin{Bmatrix} (I+\sigma_y)_A \otimes (I-\sigma_y)_B \otimes (I+\sigma_x)_C + (I+\sigma_y)_A \otimes (I+\sigma_y)_B \otimes (I-\sigma_x)_C \\ +(I-\sigma_y)_A \otimes (I-\sigma_y)_B \otimes (I-\sigma_x)_C + (I-\sigma_y)_A \otimes (I+\sigma_y)_B \otimes (I+\sigma_x)_C \end{Bmatrix} \\ + \begin{Bmatrix} (I+\sigma_y)_A \otimes (I+\sigma_x)_B \otimes (I-\sigma_y)_C + (I+\sigma_y)_A \otimes (I-\sigma_x)_B \otimes (I+\sigma_y)_C \\ (I-\sigma_y)_A \otimes (I+\sigma_x)_B \otimes (I+\sigma_y)_C + (I-\sigma_y)_A \otimes (I-\sigma_x)_B \otimes (I-\sigma_y)_C \end{Bmatrix} \\ + \begin{Bmatrix} (I-\sigma_x)_A \otimes (I-\sigma_y)_B \otimes (I-\sigma_y)_C + (I-\sigma_x)_A \otimes (I+\sigma_y)_B \otimes (I+\sigma_y)_C \\ (I+\sigma_x)_A \otimes (I-\sigma_y)_B \otimes (I+\sigma_y)_C + (I+\sigma_x)_A \otimes (I+\sigma_y)_B \otimes (I-\sigma_y)_C \end{Bmatrix} \end{bmatrix}$$

$$+\frac{p}{2}\left[(I+\sigma_z)_C \otimes (I+\sigma_z)_C \otimes (I+\sigma_z)_C + (I-\sigma_z)_C \otimes (I-\sigma_z)_C \otimes (I-\sigma_z)_C\right]$$
$$+(1-5p)(I)_A \otimes (I)_B \otimes (I)_C)$$

$$(83)$$

Therefore a sufficient condition for separability is given as

$$5p \leq 1 \quad . \quad (84)$$

An explicitly separable form for the density matrix (81) was given in [28]. The relation (84) has been discussed in various works [18, 21], showing that this condition is both sufficient and necessary for full separability.

We treat now the sufficient condition for full separability for $W$ state with probability $p$ mixed with white noise.

The W state is given by:

$$|\psi\rangle_W = \frac{1}{\sqrt{3}}\left[|0\rangle_A \otimes |0\rangle_B \otimes |1\rangle_C + |0\rangle_A \otimes |1\rangle_B \otimes |0\rangle_C + |1\rangle_A \otimes |0\rangle_B \otimes |0\rangle_C\right] \quad . \tag{85}$$

The HS decomposition of the density matrix $|\Psi\rangle\langle\Psi|_W$ is given by [1]:

$$3 \cdot 8\rho(W) = 2(\sigma_y)_A \otimes (I)_B \otimes (\sigma_y)_C + 2(I)_A \otimes (\sigma_y)_B \otimes (\sigma_y)_C + 2(\sigma_y)_A \otimes (\sigma_y)_B \otimes (I)_C +$$
$$(I+\sigma_z)_A \otimes (I+\sigma_z)_B \otimes (I-\sigma_z)_C + (I+\sigma_z)_A \otimes (I-\sigma_z)_B \otimes (I+\sigma_z)_C + (I-\sigma_z)_A \otimes (I+\sigma_z)_B \otimes (I+\sigma_z)_C$$
$$2(\sigma_x)_A \otimes (\sigma_z)_B \otimes (\sigma_x)_C + 2(\sigma_z)_A \otimes (\sigma_x)_B \otimes (\sigma_x)_C + 2(\sigma_x)_A \otimes (\sigma_x)_B \otimes (\sigma_z)_C + 2(\sigma_x)_A \otimes (\sigma_x)_B \otimes (I)_C$$
$$+2(\sigma_x)_A \otimes (I)_B \otimes (\sigma_x)_C + 2(I)_A \otimes (\sigma_x)_B \otimes (\sigma_x)_C + 2(\sigma_y)_A \otimes (\sigma_y)_B \otimes (\sigma_z)_C + 2(\sigma_y)_A \otimes (\sigma_z)_B \otimes (\sigma_y)_C$$
$$+2(\sigma_z)_A \otimes (\sigma_y)_B \otimes (\sigma_y)_C$$

$$\tag{86}$$

The $W$ state with probability $p$ mixed with white noise is given by

$$8\rho(W; mixed) = p8\rho(W) + (1-p)(I)_A \otimes (I)_B \otimes (I)_C \quad . \tag{87}$$

Then we get:

$$8\rho(W; mixed) = (1-p)(I)_A \otimes (I)_B \otimes (I)_C + +\frac{p}{3}(I+\sigma_z)_A \otimes (I+\sigma_z)_B \otimes (I-\sigma_z)_B +$$
$$\frac{p}{3}(I+\sigma_z)_A \otimes (I-\sigma_z)_B \otimes (I+\sigma_z)_B + \frac{p}{3}(I+\sigma_z)_A \otimes (I-\sigma_z)_B \otimes (I+\sigma_z)_B$$
$$+\frac{2}{3}p(\sigma_y)_A \otimes (I+\sigma_z)_B \otimes (\sigma_y)_C + \frac{2}{3}p(I+\sigma_z)_A \otimes (\sigma_y)_B \otimes (\sigma_y)_C \tag{88}$$
$$+\frac{2}{3}p(\sigma_x)_A \otimes (I+\sigma_z)_B \otimes (\sigma_x)_C + \frac{2}{3}p(I+\sigma_z)_A \otimes (\sigma_x)_B \otimes (\sigma_x)_C$$
$$+\frac{2}{3}p(\sigma_x)_A \otimes (\sigma_x)_B \otimes (I+\sigma_z)_C + \frac{2}{3}p(\sigma_y)_A \otimes (\sigma_y)_B \otimes (I+\sigma_z)_C$$

Except for the first and second row of Eq. (88), the terms in the rows 3, 4, 5 need to be written as outer products of density matrices of qubits $A, B, C$. As an example it is easy to see that

$$\frac{2p}{3}(\sigma_y)_A \otimes (I+\sigma_z)_B \otimes (\sigma_y)_C =$$
$$\frac{2p}{3}\left[\begin{array}{l}\frac{1}{2}(I+\sigma_y)_A \otimes (I+\sigma_y)_A \otimes (I+\sigma_z)_B (I+\sigma_y)_C \\ +\frac{1}{2}(I-\sigma_y)_A \otimes (I+\sigma_z)_b \otimes (I+\sigma_z)_B \otimes (I-\sigma_y)_C - (I)_A \otimes (I)_B \otimes (I)_C)\end{array}\right]. \tag{89}$$

Taking this into account, we obtain that

$$8\rho(W;mixed) = (1-5p)(I)_A \otimes (I)_B \otimes (I)_C + \sum_i (\rho_i)_A \otimes (\rho_i)_B \otimes (\rho_i)_C \quad , \quad (90)$$

where $(\rho_i)_A \otimes (\rho_i)_B \otimes (\rho_i)_C$ represent outer products of density matrices of qubits $A, B, C$, where not all $\rho_i$ are equal to the unit matrix $I$.

The sufficient condition for full separabilty of $\rho(W;mixed)$ is then obtained as:

$$5p \leq 1 \quad , \quad (91)$$

which is similar to that of GHZ mixed with white noise. By using the PT transformation, we find [1, 21] that under the condition $p > 3/(3+8\sqrt{2}) \approx 0.209589$ the density matrix of $W$ state mixed with white noise is not fully separable. We found here that under the condition $p \leq 0.2$ this density matrix is fully separable so only in a very small region the full separability problem is not clarified.

## VIII. Summary, Discussion and Conclusions

In the present work we treated the conditions for full separability and biseparability of 3-qubits MDS density matrices. We have shown that the Peres-Horodecki criterion is inconclusive for such density matrices as these density matrices and their PT have the same eigenvalues. These eigenvalues come in 4 pairs where the sum of eigenvalues for each pair is 1/4. This result can be generalized to any MDS density matrix of order odd-n where in the general case the sum of eigenvalues in one pair, is given by: $1/2^{n-1}$.

It was shown in our previous work [1] that if the sum of the absolute values of the HS parameters of the general 3-qubits state is not larger than 1, the density matrix is fully separable and we have an explicit expression for such separability. For the special case of 3-qubits MDS density matrices such sufficient condition for separability is given by Eq. (5). We proved also that any 3-qubits MDS density matrix satisfies the relation (11). The analysis of 3 qubits MDS density matrices becomes quite complicated as tensors cannot be diagonalized. We use "unfolding methods", by which tensors are converted into matrices [7-11], for treating density matrices given by the HS parameters.

A fully separable form for the 3-qubits MDS density matrix (4) related to the $l_1$ norm is given in Eq. (20). We used unfolding of the 3-qubits tensor $R_{a,b,c}$ ($a,b,c = 1,2,3$) relative to $a$ by keeping the parameter $a$ to be fixed as 1 or 2 or 3, and then the parameters $R_{1,b,c}$, or $R_{2,b,c}$, or $R_{3,b,c}$ were considered, respectively, as matrices with a $3\times 3$ dimension. Then, by using the SVD [7, 8], of $R_{1,b,c}$, and $R_{2,b,c}$, and $R_{3,b,c}$ we get for each of them 3 SV's. If the sum of these 9 SV's is not larger than 1, we conclude that the density matrix is fully separable and we have an explicitly separable form for the density matrix. Similar unfolding methods can be made, relative to $b$ or $c$. We proved in our article that the use of this procedure improves very much the condition for full separability. We demonstrated this by analyzing two examples.

A separable-like form for the 3-qubits MDS density matrix density, related to the sum of nine $l_2$ norms of triads of HS parameters was given in Eq. (44) and the corresponding sufficient condition for full separability was given by Eq. (45). We demonstrated the use of this method by analyzing the two examples treated in the previous section by the $l_1$ norm. A sufficient condition for biseparability of 3-qubits MDS density matrix, using 9 different triads of HS parameters was given in (54).

Using the HOSVD method the tensor $R_{a,b,c}$ is transformed to the tensor $S_{p,q,r}$ (Eq. (60)), yielding $\rho$ in the form (63). Using $S_{p,q,r}$ instead of $R_{a,b,c}$ we get improved conditions for full separability in section VI as demonstrated in two examples. The new Pauli matrices obtained by applying the HOSVD to $R_{a,b,c}$ ($\vec{\tilde{\sigma}}$; Eq. (63)) may of course be used for any 3-qubits density matrix (including non-MDS).

For *GHZ* and *W* states mixed with white noise with probability $p$ we found explicit fully separable forms for their density matrix showing for both cases that under the condition $5p \leq 1$ the density matrix is fully separable.